\documentclass[12pt,preprint]{aastex}

\shorttitle{Millimeter and Radio Observations of z$\sim$6 Quasars}
\shortauthors{Wang et al.}

\begin{document}


\title{Millimeter and Radio Observations of z$\sim$6 Quasars}


\author{Ran Wang }
\affil{Astronomy Department, Peking University, Beijing 100871, China}
\affil{National Radio Astronomy Observatory, PO Box 0, Socorro, NM, USA 87801}
\email{rwang@nrao.edu}
\author{Chris L. Carilli}
\affil{National Radio Astronomy Observatory, PO Box 0, Socorro, NM, USA 87801}
\email{ccarilli@aoc.nrao.edu}
\author{Alexandre Beelen}
\affil{Argelander-Institut f$\rm \ddot u$r Astronomie, University of
Bonn, Auf dem H$\rm \ddot u$gel 71, 53121 Bonn, Germany}
\affil{Institut d'Astrophysique Spatiale, Universit$\rm \acute{e}$
de Paris-Sud, F-91405 Orsay, France}
\email{abeelen@astro.uni-bonn.de}
\author{Frank Bertoldi}
\affil{Argelander-Institut f$\rm \ddot u$r Astronomie, University of
Bonn, Auf dem H$\rm \ddot u$gel 71, 53121 Bonn, Germany}
\email{bertoldi@astro.uni-bonn.de}
\author{Xiaohui Fan}
\affil{Steward Observatory, The University of Arizona, Tucson, AZ 85721}
\email{fan@as.arizona.edu}
\author{Fabian Walter}
\affil{Max-Planck-Institute for Astronomy, K$\rm \ddot o$nigsstuhl
17, 69117 Heidelberg, Germany} \email{walter@mpia.de}
\author{Karl M. Menten}
\affil{Max-Plank-Institute for Radioastronomie, Auf dem H$\rm \ddot
u$gel 71, 53121 Bonn, Germany} \email{kmenten@mpifr-bonn.mpg.de}

\author{Alain Omont}
\affil{Institut d'Astrophysique de Paris, CNRS and Universite Pierre
et Marie Curie, Paris, France} \email{omont@iap.fr}
\author{Pierre Cox}
\affil{Institute de Radioastronomie Millimetrique, St. Martin d'Heres, F-38406, France}
\email{cox@iram.fr}
\author{Michael A. Strauss}
\affil{Department of Astrophysical Sciences, Princeton University,
Princeton, NJ, USA, 08544} \email{strauss@astro.princeton.edu}
\author{Linhua Jiang}
\affil{Steward Observatory, The University of Arizona, Tucson, AZ 85721}
\email{jiang@as.arizona.edu}




\begin{abstract}

We present millimeter and radio observations of 13 SDSS quasars at
reshifts z$\sim$6. We observed eleven of them with the Max-Planck
Millimeter Bolometer Array (MAMBO-2) at the IRAM 30m-telescope at
250 GHz and all of them with the Very Large Array (VLA) at 1.4 GHz.
Four sources are detected by MAMBO-2 and six are detected by the VLA
at $\rm \gtrsim 3 \sigma$ level. These sources, together with
another 6 published in previous papers, yield a
submillimeter/millimeter and radio observed SDSS quasar sample at
z$\sim$6. We use this sample to investigate the far-infrared (FIR)
and radio properties of optically bright quasars in the early
universe. We compare this sample to lower redshift samples of
quasars observed in the submillimeter and millimeter wavelengths
((sub)mm), and find that the distribution of the FIR to B band
optical luminosity ratio ($\rm L_{FIR}$/$\rm L_{B}$) is similar from
z$\sim$2 to 6. We find a weak correlation between the FIR luminosity
($\rm L_{FIR}$) and B band optical luminosity ($\rm L_{B}$) by
including the (sub)mm observed samples at all redshifts. Some strong
(sub)mm detections in the z$\sim$6 sample have radio-to-FIR ratios
within the range defined by star forming galaxies, which suggests
possible co-eval star forming activity with the powerful AGN in
these sources. We calculate the rest frame radio to optical ratios (
$\rm R^{*}_{1.4}=L_{\nu ,1.4GHz}/L_{\nu\ 4400\AA}$) for all of the
VLA observed sources in the z$\sim$6 quasar sample. Only one radio
detection in this sample, J083643.85+005453.3, has $\rm
R^{*}_{1.4}\sim40$ and can be considered radio loud. There are no
strong radio sources ($\rm R^{*}_{1.4}\geq100$) among these SDSS
quasars at z$\sim$6. These data are consistent with, although do not
set strong constraints on, a decreasing radio-loud quasar fraction
with increasing redshift.

\end{abstract}


\keywords{galaxies: quasars --- infrared: galaxies --- radio
continuum: galaxies --- galaxies: starburst --- galaxies:
high-redshift}



\section{Introduction}
The tight correlation between the mass of supermassive black holes
(SMBH) in the centers of galaxies and the bulge mass/velocity
dispersion is well documented in the local universe (Magorrian et
al. 1998; Marconi \& Hunt 2003; Ferrarese \& Merritt 2000; Gebhardt
et al. 2000; Tremaine et al. 2002) and suggests that the evolution
of SMBH and spheroidal galaxies is coupled even at high redshift.
Due to the negative K-correction, submillimeter and millimeter
((sub)mm) observations become an efficient way to study the host
galaxies of high redshift quasars (Blain \& Longair 1993). Such
observations probe the rest frame FIR emission from the dust
components in the interstellar medium (ISM) of these objects, and
thus provide a unique chance to test co-eval black hole and bulge
formation by searching for massive starbursts in the host galaxies
of high redshift optically bright quasars.

The (sub)mm and radio properties of some optically bright quasars at
high redshift are discussed in a number of papers. Observations of
large samples of optically selected quasars from z$\sim$1.5 to 5
(Omont 2001, 2003; Carilli et al. 2001; Isaak et al. 2002; Priddey
et al. 2003a; Beelen et al. 2004) result in a (sub)mm detection rate
of about 30\% at mJy sensitivity. The derived far-infrared (FIR)
luminosities of the (sub)mm detections are typically $\sim10^{13}\rm
\,L_{\odot}$ and imply dust masses $\geq10^{8}\rm \, M_{\odot}$.
Comparing samples observed with the Max-Planck Millimeter Bolometer
(MAMBO) at the 30-meter IRAM telescope at redshift 2 and 4, Omont et
al. (2003) found that the FIR luminosities of the optically bright
quasars in these samples do not evolve with redshift. Statistical
tests for these (sub)mm observed quasars show a weak correlation
between the optical and FIR luminosities (Omont et al. 2003; Beelen
2004; Cox et al. 2005), which is argued as an evidence for FIR
emission from warm dust heated by star formation. The inferred star
formation rate is $\sim10^{3}\rm \, M_{\odot}\ yr^{-1}$ when
assuming a normal initial mass function (IMF). This interpretation
is supported by the fact that the radio-to-FIR ratios for the
FIR-luminous sources are consistent with the range spanned by star
forming galaxies (Carilli et al. 2001; Petric et al. 2006).

The Sloan Digital Sky Survey (SDSS, York et al. 2000) has identified
19 bright quasars at z$\sim$6 (Fan et al. 2000, 2001, 2003, 2004,
2006a). These quasars are unique, in that they are undergoing rapid
accretion onto SMBHs with masses $\gtrsim 10^{9}\rm \,M_{\odot}$,
within 1 Gyr of the Big Bang, an epoch approaching cosmic
reionization (Fan et al. 2006c). These z$\sim$6 quasars have been
observed at all wavelengths from the X-ray to the radio (eg.
Bechtold et al. 2003; Pentericci et al. 2003; Shemmer et al. 2006;
Jiang et al. 2006a; Petric et al. 2003). The X-ray to near-infrared
(NIR) observations imply Spectral Energy distributions (SEDs)
similar to those of local quasars. The integrated bolometric
luminosities exceed $\gtrsim10^{13}\rm \, L_{\odot}$, and the black
hole masses are $\gtrsim10^{9}\rm \, M_{\odot}$ (Willott et al.
2003; Iwamuro et al. 2004; Jiang et al. 2006a). Seven z$\sim$6 SDSS
quasars (Fan et al. 2000, 2001, 2003) were observed at 350 GHz with
the SCUBA camera at the James Clerk Maxwell Telescope or at 250 GHz
with MAMBO. Three are detected by SCUBA (Priddey et al. 2003b;
Robson et al. 2004) and two are detected by MAMBO (Bertoldi et al.
2003a). Deep VLA observations at 1.4 GHz were made for six sources,
resulting in detections of two of them (Petric et al. 2003; Carilli
et al. 2004).


SDSS J114816.64+525150.3 at $z=6.42$ is the best studied FIR
luminous source at z$\sim$6. It was detected with SHARC II, the
bolometer camera at the Caltech Submillimeter Observatory at 350
$\rm \mu$m, SCUBA at 450 $\rm \mu$m and 850 $\rm \mu$m, MAMBO at 1.2
mm, and the VLA at 1.4GHz (Beelen et al. 2006; Robson et al. 2004;
Bertoldi et al. 2003a; Carilli et al. 2004). Fits to the SED from
the FIR to radio imply a warm dust component with a temperature of
55K (Beelen et al. 2006). The corresponding FIR luminosity is
$2.2\times10^{13}\rm \, L_{\odot}$ with an inferred dust mass of
$4.5\times10^{8}\rm \, M_{\odot}$. It is also the only z$\sim$6
source that has been detected in the CO and [C II] 158 $\rm \mu$m
emission lines (Bertoldi et al. 2003b; Walter et al. 2003; Walter et
al. 2004; Maiolino et al. 2005). These CO observations reveal a
large mass of molecular gas ($\sim2\times10^{10}\rm \, M_{\odot}$)
in the host galaxy.

The (sub)mm through radio observations of J114816.64+525150.3 argue
for a massive starburst in its host galaxy, because (i) the FIR to
radio ratio of this source follows that of typical star forming
galaxies in the local universe (Yun et al. 2001; Carilli et al.
2004), (ii) the huge amount of molecular gas in its host galaxy can
provide the required fuel for massive star formation, and (iii) the
star formation rates indicated by both FIR luminosity and [C II]
158$\mu$m line luminosity are $\sim 10^3$ \rm M$_\odot$ yr$^{-1}$
(Bertoldi et al. 2003a; Maiolino et al. 2005).

Nineteen z$\sim$6 SDSS quasars have been published to date (Fan et
al. 2006b). These are an optically selected sample at the highest
redshift. We are pursuing a series of (sub)mm through radio studies
on this sample in order to (i) find the general dust and gas
properties in the host galaxies of these highest redshift quasars,
and (ii) search for star formation activity co-eval with the rapid
growth of SMBHs in the early universe.

In this paper, we present new MAMBO-2 250 GHz observations of
eleven, and VLA 1.4GHz observations of thirteen, z$\sim$6 SDSS
quasars. Then, together with previously published results, we
discuss the general far-infrared and radio properties of the
optically selected quasars at z$\sim$6, focusing on luminosity
correlations and evolution. We will also discuss the radio loud
fraction of quasars. We will present further analysis of the FIR to
radio SEDs and discuss possible star forming activity in a second
paper (paper II). The sample and observations are described in
section 2, and results are presented in section 3. In section 4 and
5, we analyse and discuss the general properties of FIR and radio
luminosities for the entire sample and give the conclusion in
section 6. We adopt a concordance cosmology with $H_{0}=71\rm\, km\,
s^{-1}\, Mpc^{-1}$, $\Omega_{m}=0.27$ and $\Omega_{\lambda}=0.73$
throughout this paper.


\section{Sample and Observation}



The sample of z$\sim$6 quasars are selected from about 6600 $\rm
deg^2$ of imaging data from the Sloan Digital Sky Survey (Fan et al.
2006b). Fan et al. (2000, 2001, 2003, 2004, 2006a) selected sources
with very red $i-z\, colors$ as z$\geq5.7$ quasar candidates. These
candidate sources were followed up with high quality spectroscopy.
Nineteen z$\sim$6 quasars have been published by the SDSS survey
(Fan et al. 2000, 2001, 2003, 2004, 2006a) with redshifts ranging
from 5.74 to 6.42 and rest frame B band optical luminosities\footnote{$\rm
L_{B}\equiv \nu L_{\nu,4400\AA}$: $L_{\nu,4400\AA}$ is the luminosity
density at rest frame $\rm 4400\AA$. we calculate $L_{\nu,4400\AA}$ and $L_{B}$
from the AB magnitude at rest frame $1450\AA$ in the discovering papers 
(Fan et al. 2000, 2001, 2003, 2004, 2006a), assuming
an optical spectral index of -0.5. } from $10^{12.5} \rm\,
L_{\odot}$ to $10^{13.3} \rm\, L_{\odot}$. Most of the 13 objects we
observed in this work were discovered in the past two years (Fan et
al. 2004; 2006a) and have not been observed in the (sub)mm or radio
bands before.

We compare the (sub)mm properties of the z$\sim$6 objects with
luminous (\rm $L_{B}\ge10^{12.5}\rm\, L_{\odot}$) quasars at lower
redshifts from the literature. We define a low redshift group, $\rm
1.5\le z \le3.0$, using a sample from Omont et al. (2003) containing
radio-quiet quasars with B-band absolute magnitude $\rm -29.5\le
M_{B}\le-27.0$ and redshift $\rm 1.8\le z \le2.8$ observed by MAMBO
at 250GHz, and a sample from Priddey et al. (2003a) of radio-quiet
quasars with $\rm -29.2\le M_{B}\le-27.5$ and $\rm 1.5\le z \le3.0$.
Our higher redshift group ($\rm 3.6\le z \le5.0$) is drawn from the
the Palomar Sky Survey (PSS) selected sample of Omont et al. (2001)
($\rm M_B\le-27.0$; $\rm 3.9\le z \le4.6$) and the SDSS sample of
Carilli et al. (2001) ($\rm -28.8\le M_B\le-26.1$; $\rm 3.6\le z
\le5.0$), both observed with MAMBO at 250 GHz. The 3$\rm \sigma$
detection limits of the MAMBO observations by Omont et al. (2001)
and (2003) are $\sim$1.5 to 4 mJy, and $\sim$1.4 mJy by Carilli et
al. (2001). The typical 3$\rm \sigma$ detection limit of SCUBA
observations reported by Priddey et al. (2003a) is $\rm \sim10$ mJy
at 350 GHz, which corresponds to an upper limit of $\rm \sim$4 mJy
at 250 GHz (assuming a warm dust SED with temperature $\rm
T_{d}=47K$ and emissivity index $\rm \beta=1.6$).




A total of 13 z$\sim$6 SDSS quasars were observed in the course of
the work presented here, all with the VLA at 1.4GHz and eleven (all
but J000552.34-000655.8 and J104845.05+463718.3) with MAMBO-2 at
250GHz (See Table 1 for the source list). J104845.05+463718.3 was a
published detection by MAMBO (Bertoldi et al. 2003a) and was
observed by the VLA with a signal at 2.2$\rm \sigma$ (Carilli et al.
2004). We re-observed it with the VLA, and combined all the data.

MAMBO-2 at the IRAM 30m telescope is a 117-channel bolometer array.
The Half Power Beamwidth (HPBW) of each pixel is $\rm 11''$ and the
spacing between horns is about $\rm 20''$. The effective sensitivity
is about 40 mJy$\rm s^{1/2}$. The new observations were made in the
winters of 2004-2005 and 2005-2006 during pooled observations using
the on-off mode, wobbling by $\rm 32''-46''$ in azimuth and at a
rate of 2 Hz. The target sources are positioned on the most
sensitive bolometer and the correlated sky noise is determined from
the other bolometers and subtracted from the source bolometer. The
median rms is $\sim$0.8 mJy (Table 1).
We reduced the data with the MOPSIC software package (Zylka 1998)
and standard scripts for on-off observation data.

The VLA observations were made with the A array, of which the 
maximum baseline is 30 km. The sources were observed at 1.4 GHz with
two Intermediate Frequency bands "IFs" and 50MHz bandwidth per IF.
The corresponding Full Width at Half Maximum (FWHM) resolution is
about $\rm 1.4''$. We observed each source for two to four hours to
a typical rms noise level of $\sim$16 $\rm \mu Jy\,beam^{-1}$ (Table
1). The data were reduced and images were made using the standard
VLA wide field data reduction software AIPS.




\section{Results}

We present our MAMBO-2 results of 11 sources and the VLA results of
13 sources in Table 1. The optical properties are taken from Fan et
al. (2006b) and presented in Col. (1): the SDSS name, Col. (2): the
redshift, and Col. (3): the AB magnitude at 1450$\rm \AA$. Col. (4)
lists the radio 1.4GHz surface brightness at the optical position.
Col. (5), (6), and (7) list the nearest radio peak position and peak
surface brightness for detected sources. Col. (8) presents the
250GHz flux densities in mJy. We mark detections in bold face. For
the source J000552.34-000655.8, there is a strong ($\rm\sim$Jy) radio 
source in the field which precludes deep VLA radio imaging. The rms on the  
1.4GHz map is 130 $\mu$Jy, which is an order of magnitude higher than the
others, yielding an extremely high upper limit of $\sim$390 $\mu$Jy.
Thus, we exclude this source in all of our analyses in this
paper.

We show the radio 1.4GHz continuum images of the 12 remaining sources in
Figure 1. We searched for $\rm \geq3\sigma$ peaks within a $\rm 0^{''}.6$
radius from the optical quasar position in the radio map. This is a
combination of the positional uncertainty of $\rm 0^{''}.3$ between the
radio and optical reference frames (Deutsch 1999) and the radio
observation uncertainty of $\rm \sigma
_{pos}\sim\frac{FWHM}{SNR}\sim0^{''}.5$\footnote{We require a signal
to noise ratio (SNR) of 3 for detections here.}. Six sources are
detected at the $\rm \geq3\sigma$ level. According to the 1.4 GHz radio
source counts (Fomalont et al. 2006, in prep.), we expect 0.003
detections with $\rm S_{1.4GHz}\geq50\,\mu Jy$ by chance, within the
total search area around the 12 quasars.

We obtain four detections among the 11 sources with MAMBO. Three of
the detections, J084035.09+562419.9 J092721.82+200123.7 and
J133550.81+353315.8, are detected at $\rm >4\sigma$. The fourth
source, J081827.40+172251.8, is marginally detected at the $\rm
\sim3 \sigma$ level.
As a comparison, the cumulative
number counts of 250 GHz source is about 400 deg$^{-2}$ with flux
density $\rm S_{250GHz}\geq 2.4\,mJy$ (Greve et al. 2004; Voss et al. 2006) 
which is the typical 3$\rm \sigma$ limit of our observations.
Thus we should expect 0.03 detections in our 11 fields by chance,
given the $\rm 11''$ beam size of the MAMBO bolometer elements.

We summarize the (sub)mm and radio observations of the six
previously observed quasars in the z$\sim$6 SDSS sample (Petric et
al. 2003; Priddey et al. 2003b; Bertoldi et al. 2003a; Robson et al.
2004; Carilli et al. 2004) in Table 2. Col. (1), (2), and (3) give
the name, redshift, and 1450$\rm \AA$ AB magnitude, and Col. (4),
(5), (6), and (7) the flux densities at 250 GHz, 350 GHz, 667 GHz
and 1.4 GHz.

Combining Tables 1 and 2, there are seventeen z$\sim$6 quasars that
have been observed with MAMBO. Together with the results from SCUBA
at 850 $\mu$m, there are eighteen z$\sim$6 SDSS quasars that have
(sub)mm observations at the 1 mJy sensitivity level and 8 of them
are detected at $\rm \gtrsim 3\sigma$. The detection rate is $\rm
44\pm16\%$. The detection rate of optically bright quasar samples at
redshifts 2 and 4 is $\rm \sim30\%$ in Omont et al. (2001, 2003)
with an rms level of $>1.5\,mJy$, and $\rm \sim39\%$ in Carilli et
al. (2001) with a lower rms level of $\sim 1.4\,mJy$. Thus our
detection rate at z$\sim$6 is slightly higher compared to these
observations at lower redshifts, but is consistent within the
errors.

Seventeen sources in the z$\sim$6 SDSS quasar sample have deep
observations with the VLA at 1.4 GHz and 8 are detected at the $\rm
\gtrsim 3\sigma$ level. J083643.85+005453.3 is the only $>$1 mJy
radio source at 1.4 GHz among them. It was detected by Petric et al.
(2003), as well as in the FIRST survey (White et al. 1997), leading
to a weighted average flux density of 1.74$\pm$0.04 mJy\footnote{The
1.4 GHz flux density for J083643.85+005453.3 is $\rm 1750\pm40\,\mu Jy$
in Petric et al. (2003) and $\rm 1530\pm150\,\mu Jy$ in the FIRST
catalog. }. Another source, J163033.90+401209.6, has not been
observed with the VLA yet, and the FIRST catalog yields an upper
limit of 0.44 mJy at the optical position (Bertoldi et al. 2003). We
thus exclude this source in the analysis of radio properties in the
next section.

\section{Analysis}

\subsection{The FIR emission}

To estimate the FIR luminosity of our z$\sim$6 quasars, we model
the FIR continuum with an optically thin grey body (Eq. (1)), adopting a dust
temperature of $\rm T_{d}=47K$ and emissivity index, $\rm
\beta=1.6$, which is derived from the mean SED of high-redshift
quasars (Beelen et al. 2006). We normalize the SED to the (sub)mm
data and integrate the model SED from rest frame 42.5 $\rm \mu m$ to
122.5 $\rm \mu m$ to get the total FIR flux\footnote{For
J114816.64+525150.3, we directly adopt the fitting results from
Beelen et al. (2006) with $\rm T_{d}=55K$ and
$\rm \beta=1.6$.}:\\
\begin{equation}
\rm FIR\propto\int \nu ^{3+\beta}\frac{1}{exp(h\nu /kT_{d})-1}d\nu\\
\end{equation}
This estimation of FIR emission is sensitive to the assumed dust
temperature ($\rm T_{d}$). A higher temperature, eg. $\rm
T_{d}=55K$, will increase the estimated FIR flux by a factor of
$\sim$1.5. For non-detections ($<\,3\sigma$), we adopt as upper
limits the larger value of either (a) the 2$\rm \sigma$ rms or (b)
the measured value at the optical position plus 1$\rm \sigma$ rms
for all the following calculations and plots. The FIR luminosities
($\rm L_{FIR}$) are given in table 4; most of the (sub)mm detected
sources are very luminous in the FIR band, with implied $\rm
L_{FIR}\sim10^{13}\, L_{\odot}$ (see Table 4). We estimated the FIR
luminosities for all the low redshift comparison samples in the same
way.

We plot the FIR luminosity ($\rm L_{FIR}$) versus redshift in Figure 2, including
the z$\sim$6 SDSS sample and all the comparison samples. In Figure
3, we present the histograms of the $\rm L_{FIR}$ and $\rm
L_{FIR}/L_{B}$ distributions separately for different redshift
intervals. For the (sub)mm detections at all redshifts, the $\rm
L_{FIR}$ values lie between $10^{12.5}\rm\, L_{\odot}$ and
$10^{13.8}\rm\, L_{\odot}$. The $\rm 1.5\le z \le 3$ group covers an
even narrower range with most of the detected sources distributed at
$\rm L_{FIR}\ge10^{13}\, L_{\odot}$ These ranges are partially a
result of the detection thresholds, i.e. a detection limit of 1.5mJy
at 250GHz corresponds to FIR luminosities $\sim10^{12.5} \rm\,
L_{\odot}$ for redshifts from 4 to 6 and $\sim10^{13} \rm\,
L_{\odot}$ at z$\sim$2. However, the $\rm\, L_{FIR}/L_{B}$
distributions are similar for the (sub)mm detections at all
redshifts (See Fig. 3 (b)). A similar conclusion was
obtained by Omont et al. (2003) and Beelen et al. (2004), but we are
extending it to the quasar sample at z$\sim$6. This fact suggests
the optical to FIR spectra energy distributions (SEDs) of the
luminous quasars in both optical and FIR bands do not evolve much
from z$\sim$2 to 6.

$\rm L_{FIR}$ versus $\rm L_{B}$ is plotted in Figure 4, for quasars
at all redshifts. We present correlation tests for (i) the z$\sim$6
sample only (z$\sim$6), (ii) samples at all redshifts (Total), and 
(iii) (sub)mm detections in all samples (Detection). We apply the
General Kendall's tau test (Isobe et al. 1986), taking $\rm L_{B}$
as the independent variable and $\rm L_{FIR}$ as the dependent
variable. The results are listed in Table 3 with Col. (1) the sample
used, Col. (2) the sample size, and Col. (3) the Kendall's tau
value. The probabilities ($\rm P_{null}$) that no correlation exists
between $\rm L_{B}$ and $\rm L_{FIR}$ (the null hypothesis) are
given in Col. (4). We take $\rm P_{null}=5\%$ as the significance
level below which the null hypothesis can be rejected.

There is no correlation between $\rm L_{B}$ and $\rm L_{FIR}$ for
the z$\sim$6 sample only ($\rm P_{null}\sim44\%$). 
However, the $\rm P_{null}$ value is 6\% when all the samples 
are included, which suggests a marginal correlation. 
These results indicates that the FIR and optical emission of
these (sub)mm observed quasars are correlated, but the scatter is
large enough that the correlation is not seen over a narrow
luminosity range. This correlation is even more significant 
when we do the test with only the (sub)mm detections at all 
redshifts ($\rm P_{null}\sim0.02\%$). But one should be 
careful with this result as there may be a number of 
observational or selection bialses in the sample of detections. 
For example, given the similar observational sensitivity level, 
the FIR luminosity threshold for the (sub)mm detections
is decreasing with redshifts at $z>2$ (see Figure 1).
We apply linear regression to all of the (sub)mm
observed samples (Total) and the subsample of detections
(Detection), using the Expectation-Maximization (EM) method which
can deal with data including upper limits (Isobe et al. 1986). The
best
fitting results are:\\
\begin{equation}
\rm Total:\ log\,(\frac{L_{FIR}}{L_{\odot}})=(0.21\pm0.15)log(\frac{L_{B}}{L_{\odot}})+(9.67\pm2.03)\\
\end{equation}
\begin{equation}
\rm Detection:\ log\,(\frac{L_{FIR}}{L_{\odot}})=(0.40\pm0.09)log(\frac{L_{B}}{L_{\odot}})+(7.82\pm1.20)\\
\end{equation}

The regression results show a non-linear
relationship between the FIR emission and the quasars'
optical emission with slopes much smaller than unity. 
We plot the fitting results in Figure 4. As a comparison,
We calculate the typical FIR-to-B band luminosity ratios
for local optical quasars based on the radio quiet and radio
loud quasar templates in Elvis et al. (1994). The FIR
luminosities are integrated directly from the template SEDs.
The derived FIR-to-B band luminosity ratios ($\rm L_{FIR}/L_{B}$)
are 0.29 and 0.38 for the radio quiet and radio loud templates
respectively, and are ploted as $\rm L_{FIR}=0.29L_{B}$ (solid line)
and $\rm L_{FIR}=0.38L_{B}$ (long dashed line) in Figure 4.
According to Figure 4, most of the (sub)mm detections have larger $\rm L_{FIR}/L_{B}$ 
value than that of the local optical quasar templates. For the eight 
(sub)mm detected quasars at z$\sim$6, six of them have FIR emission 
stronger than that predicted from the radio quiet template (by factors 
from $\rm \sim1.5$ to $\rm \gtrsim5$), and only one source J081827.40+172251.8 
is consistent with the template within errors.

\subsection{The Radio emission}

The rest frame 1.4GHz luminosities ($\rm L_{Rad}$\footnote{$\rm
L_{Rad}\equiv \nu L_{\nu,1.4GHz}$ in rest frame}) for the 17 quasars observed with
the VLA in the z$\sim$6 sample are calculated, assuming a power law
($\rm f_{\nu}\sim \nu ^{\alpha}$) radio SED and a radio spectral
index of $\rm \alpha = -0.75$ (Condon 1992). Since all the sources
are unresolved on the radio maps (Figure 2), we adopt the peak
surface brightness in Table 1 as the 1.4GHz flux density for our VLA
detections and take the larger value of either the $\rm 2\sigma$ rms
on the radio map or the measured value at the optical position plus
1$\rm \sigma$ rms as the upper limits for non-detections. The rest
frame radio to B band optical luminosity density ratios $\rm
R^{*}_{1.4}=L_{\nu ,1.4GHz}/L_{\nu,4400\AA}$ (Cirasuolo et al. 2003;
2006) are calculated and listed in Table 4. $\rm L_{\nu,4400\AA}$ is
the luminosity density at rest frame $\rm 4400\AA$ (See the footnote
in Section 2). We adopt $\rm
R=L_{\nu,5GHz}/L_{\nu,4400\AA}=10$ (Kellermann et al. 1989) to
separate radio loud and radio quiet sources. This corresponds to
$\rm R^{*}_{1.4}\sim30$ by assuming the above radio spectral index
and converting the rest frame 5 GHz flux density to 1.4 GHz.

We plot the rest frame 1.4 GHz radio luminosity ($\rm L_{Rad}$) vs. B
band luminosity ($\rm L_{B}$) in Figure 5. The typical 3$\rm \sigma$
detection limit of the VLA observation is $\rm \sim50\,\mu$Jy (dotted
line), which is roughly 10 times deeper than the FIRST survey (solid
line). Thus our VLA observations are sensitive enough to detect any radio loud source
with B band luminosity $\rm >10^{12}\,L_{\odot}$ at z$\sim$6 for all of
the 17 sources. We plot the $\rm R^{*}_{1.4}$ distribution of the 17
z$\sim$6 quasars in Figure 6. None of the 13 sources with new radio
data reported in this paper is radio-loud. For the whole sample of 17
z$\sim$6 quasars, there is one marginally radio-loud object,
J083643.85+005453.3, with $\rm R^{*}_{1.4}\sim40$.

The deep VLA observations also enable us to investigate the
correlation between radio and FIR emission of the z$\sim$6 quasars.
Yun et al. (2001) employ a $q$ parameter
to quantify the ratio of the FIR and radio luminosities:\\
\begin{equation}
\rm q\equiv log(\frac{FIR}{3.75\times10^{12}\,W\,
m^{-2}})-log(\frac{f_{1.4GHz}}{W\, m^{-2}\,Hz^{-1}}).\\
\end{equation}
Yun et al. (2001) studied the infrared-selected galaxies in the IRAS
2Jy sample and found a $q$ value of 2.34 for typical star forming
galaxies. We calculate the $\rm q$ values for the 12 z$\sim$6
quasars that are detected at either radio or (sub)mm wavelengths.

We also plot $\rm L_{FIR}$ vs. the rest frame 1.4GHz radio
luminosity density ($\rm L_{\nu,1.4GHz}$) in Figure 6, comparing the
z$\sim$6 quasar sample to typical star forming galaxies in the local
universe (Yun et al. 2001). The solid line in Figure 6 corresponds
to the typical $\rm q$ value of $2.34$ in star forming galaxies, and
the dotted lines denote excesses that are 5 times above and below
this typical correlation (Yun et al. 2001). 

There are four sources in the z$\sim$6 quasar sample that are
detected in both the radio and (sub)mm regimes. One of them,
J081827.40+172251.8, has a small q value that falling
beyond the range defined by typical star forming galaxies,
indicating the dominance of AGN power in the FIR to radio
SED of this source. However, the FIR-to-radio ratios of the other 
three sources, though slightly above the mean value of $\rm q=2.34$, 
are consistent with star forming galaxies (Figure 6 and Table 4).


\section{Discussion}

The heating sources of the FIR emitting dust in quasars
are studied and the contributions from host galaxy star formation
are discussed in a number of papers (eg. Haas et al. 2003; Omont et al. 2003;
Schweitzer et al. 2006). Star formation is believed to be
the dominated dust heating resource in the local infrared luminous
quasars that are located in ULIRGs (Hao et al. 2005). Schweitzer et
al. (2006) studied the connection between FIR emission and PAH/low
excitation fine-structure emission lines of the local PG quasars.
Their results suggest that star formation is responsible for at
least 30\% to the average local quasar FIR emission and that this
contribution tends to increase with FIR luminosity. The z$\sim$6
SDSS quasars we discussed in this paper belong to the population of
the brightest quasars in the optical. About $\rm 1/3$ of them are detected
at (sub)mm wavelength. According to Figure 4, most of these
detections have FIR luminosities stronger than the predictions from
local quasar templates, indicating extra
emission from warm dust in these sources compared to the
typical optical quasar emission. Moreover, the relationship
between the FIR and B band optical emission is non-linear 
and significantly scattered, only manifesting itself when
a larger luminosity range is considered (See Section 4.1).
Beelen (2004) studied a number of optically bright quasar samples at
$\rm z>1.5$ that had (sub)mm observations, and first report this weak
correlation between FIR and B band optical emission,
namely $\rm logL_{FIR}=(8.36\pm 0.90)+(0.33\pm 0.07)\times logL_{B}$
(See also Cox et al. 2005). The results are consistent when we
include our new observations at z$\sim$6. Additionally,
most of the (sub)mm detections at z$\sim$6 have FIR-to-radio ratios
or upper limits within the range occupied by star forming
galaxies (See Table 4 and Figure 6).

One possible explanation for all of these facts
is that star formation is happening on a massive scale in the host
galaxies of these FIR luminous z$\sim$6 quasars and
dominating the heating process of the FIR emitting warm dust.
The implied star formation rate
is $\rm \sim 10^3\,M_{\odot}\,yr^{-1}$. 
This idea is supported by the detections of CO in the 
high-redshift quasars that have strong FIR 
emission ($\rm L_{FIR}\gtrsim10^{13}\,L_{\odot}$). 
CO line emission was detected in several FIR luminous 
quasars at z$\gtrsim$4, including the highest redshift 
quasar J114816.64+525150.2 (eg. Riechers et al. 2006; 
Bertoldi et al. 2003b; Walter et al. 2003, 2004). 
The CO detections indicate huge amount of molecular 
gas in the host galaxies of these FIR luminous 
quasars which are the requisite fuel for star 
formation. Moreover, Riechers et al. (2006) found 
that the FIR-to-CO relation of the high-redshift 
FIR luminous quasars are consistent with that defined by  
ULIRGs, starforming galaxies and submillimeter 
galaxies. 

One may argue that the strong FIR emission can be also
processed by the AGN power since the dust torus geometry 
of these sources are still unkown. This possibility 
cannot be rule out given the limited data we have 
for these sources, but detailed dust models are required 
to explain the results we list above. One fact is that the hot 
dust emission of the SDSS z$\sim$6 quasars probed by recent  
Spitzer observation are similar to those of the local quasar templates 
as we mentioned in Section 1 (Jiang et al. 2006a). 
Four (sub)mm detections are included in this Spitzer 
observed sample, and no different properties are found in the 
optical to near infrared SEDs of these sources. This may 
not support the idea of a quite different geometry of the 
dust torus.

However, we should mention that there are still large
uncertainties in the estimations of the FIR luminosities
for these (sub)mm detected z$\sim$6 quasars.
The FIR luminosities for some of the sources can be
overestimated given the poor data at (sub)mm wavelengths. 
Thus further studies on these (sub)mm detected z$\sim$6 quasars are
required, including observations of the submm continuum
at higher frequencies and emission lines of CO and other
molecules, such as HCN, as well as the C and $\rm C^+$
fine structure lines. 

Another interesting question is how the current radio observations
constrain the radio loud fraction (RLF) at z$\sim$6. The RLF of
optically selected quasar samples is $\rm \sim10\%$, as quoted in
many papers (Kellermann et al. 1989; Ivezi$\rm \acute{c}$ et al.
2002, 2004; Cirasuolo et al. 2003, 2006).
Jiang et al. (2006b) studied the FIRST data (Becker et al. 1995) for
SDSS quasars with redshifts up to 5, and suggested that the quasar
RLF increases with optical luminosity and decreases with redshift,
namely $\rm
log\,((RLF)/(1-RLF))\,=\,(-0.112\pm0.109)+(-2.196\pm0.269)log(1+z)+(-0.203\pm0.026)(M_{2500}+26)$.
This result gives an RLF of $\rm \sim37\%$ at $\rm z=0.5$, $\rm \sim11\%$ at $\rm z=2$, and $\rm
\sim2\%$ at $\rm z=6$, given the rest frame 2500$\rm \AA$ absolute
AB magnitude $\rm M_{2500}=-27.3$
--- the typical value of the z$\sim$6 SDSS quasar 
sample\footnote{The $\rm M_{2500}$ is estimated from the absolute
AB magnitude at rest frame $\rm 1450\AA$ ($\rm M_{1450}$, Fan et al. 2006b). For the
current sample of 19 z$\sim$6 SDSS quasars with $\rm M_{1450}$
from -26.2 to -27.9, we adopt a typical value of $\rm M_{1450}=-27.0$
and an optical spectral power law index of -0.5.}.

There is one marginally radio loud source in this z$\sim$6 sample of
17 sources. This result argues against a high RLF, i.e. $>20\%$, 
for the current SDSS quasar sample at z$\sim$6. Thus it is roughly consistent 
with the result of Jiang et al. (2006b). However, the
sample size is still too small to set a strict constrain, 
i.e. differentiate an RLF between $\rm \sim2\%$ and $\rm \sim10\%$. 
Thus to provide a better test of the redshift evolution 
of the RLF; we would need a sample three times
larger to usefully constrain the RLF at z$\sim$6.

We should also mention the recent discovery of a radio loud quasar
at z$>$6, FIRST J1427385+331241(McGreer et al. 2006), with a FIRST
flux density of 1.73mJy. This source is the most radio luminous
quasar known at z$\sim$6, with $\rm R^{*}_{1.4}>100$. However, it
cannot be included in our radio loud analysis at z$\sim$6, as the
selection criteria of this source are quite different from that of
the SDSS quasars and its UV/Optical emission is just beyond the SDSS
detection limits (McGreer et al. 2006).

\section{Conclusion}


In this paper, we present new results of millimeter and radio
observations of a sample of z$\sim$6 quasars. These quasars are
selected from the SDSS survey and observed with MAMBO-2 at 250 GHz
and the VLA at 1.4 GHz. We obtained three $\rm >4\sigma$ detections
and one $\rm \sim3\sigma$ detection out of 11 observed sources by
MAMBO-2 and six radio detections out of 13 observed sources by the
VLA.

We combine our new millimeter and radio results of the
SDSS z$\sim$6 quasars with results from the literature and discuss the
FIR and radio properties of the optically selected quasars at
z$\sim$6.
Our conclusions are as follows.\\
\vspace{-2.5em}
\begin{itemize}
\setlength{\itemsep}{-2.0em}
\item Eight out of
18 z$\sim$6 optically selected quasars are detected in the (sub)mm
regime. This indicates a (sub)mm detection rate of $44\%$ at
z$\sim$6 at mJy sensitivity. Within the errors, this is consistent
with the $\sim$30\% (sub)mm detection rate at lower reshift (eg.
Carilli et al. 2001; Omont et al. 2001;
2003). The observational data imply FIR luminosities $\rm \sim10^{13}\rm\, L_{\odot}$ in the (sub)mm detected sources.\\
\item We compare the distribution of FIR luminosities
and FIR to optical ratios between the z$\sim$6 SDSS sample and
(sub)mm observed optically bright quasars ($\rm
L_{B}\ge10^{12.5}\rm\, L_{\odot}$) at lower reshifts. The
distributions of the FIR-optical ratio are similar for different
redshift groups, which suggests that the average optical-to-FIR
SED of optically bright quasars is independent of redshift.\\
\item We extend the quasar FIR-to-optical correlation
study to the z$\sim$6 SDSS sample. No correlation is found with the
z$\sim$6 sample only. However, a correlation (albeit with large
scatter) can be seen when all the samples extending from z=1.5 to
6.42 are included.\\
\item We also discussed the FIR-to-radio ratios of
the z$\sim$6 quasars, by comparing them to the typical correlation
defined by star forming galaxies. Three of the four sources that are
detected in both the millimeter and radio bands have FIR-to-radio
ratios within the range
defined by star forming galaxies.\\
\item We found no strong radio sources with $\rm R^{*}_{1.4}\geq30$
among the new SDSS sources of bright z$\sim$6 quasars. In the whole
z$\sim$6 sample, only one radio detection has a radio to optical
ratio $R^{*}_{1.4}\sim40$ and no source has $R^{*}_{1.4} \gtrsim
100$. These data are consistent with, although do not set strong
constraints on, the recent conclusion of a decreasing
radio loud quasar fraction with increasing redshift (Jiang et al. 2006b).\\
\end{itemize}
\vspace{-2.5em}

These
results give a view of the general FIR through radio properties of
the z$\sim$6 SDSS quasars. The data are consistent with the idea
that massive starbursts may exist in the host galaxies of the strong
(sub)mm detections at z$\sim$6 and contribute to the FIR and radio
emission.
These strong (sub)mm sources provide
the only candidates to search for CO and $\rm C^{+}$ into the epoch
of reionization, and to test the idea of co-eval SMBH and host
galaxy formation. We may expect to go an order of magnitude deeper
in a few years time with the coming instruments of the Expanded Very
Large Array (EVLA) and the Atacama Large Millimeter Array (ALMA).



\acknowledgments We acknowledge support from the Max-Planck Society
and the Alexander von Humboldt Foundation through the
Max-Planck-Forschungspreis 2005. The National Radio Astronomy
Observatory is a facility of the National Science Foundation,
operated by Associated Universities, Inc. We thank Dr. R. Zylka at
IRAM for his help with the MAMBO data reduction. Ran Wang thanks her
supervisor, Prof. Xue-Bing Wu and the whole AGN group at Peking
University for helpful discussions. X. Fan acknowledge support from
NSF grant AST 03-07384, a Sloan Research Fellowship, and a Packard
Fellowship for Science and Engineering.



{\it Facilities:} \facility{VLA}, \facility{IRAM (MAMBO)},
\facility{SDSS}.


\begin{figure}
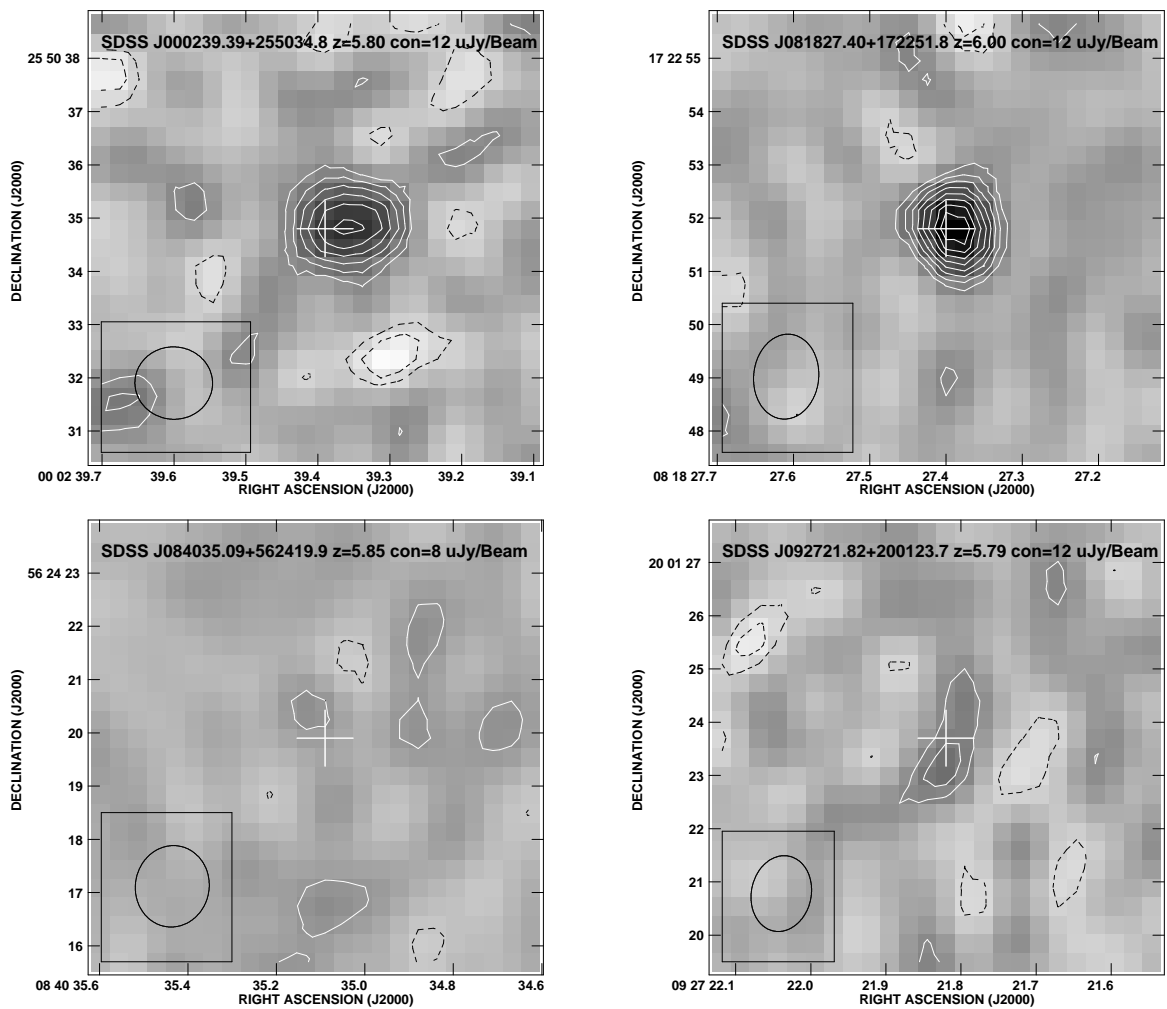

\plottwo{f1a.eps}{f1b.eps}\\
\plottwo{f1c.eps}{f1d.eps}\\
\figcaption{VLA images of 12 z$\sim$6 quasars at 1.4GHz at 1.4$''$
resolution (FWHM). The parameter 'con' denotes the value for the
contour increment in each map in units of $\rm \mu Jy/Beam$. The
contour levels are (-3, -2, 2, 3, 4, 5, 6, 7, 8, 9, and 10)
$\times$con. The ellipse indicates the beam in each case, and the
cross marks the optical position. \label{fig1}}
\end{figure}

\begin{figure}
\figurenum{1}
\plottwo{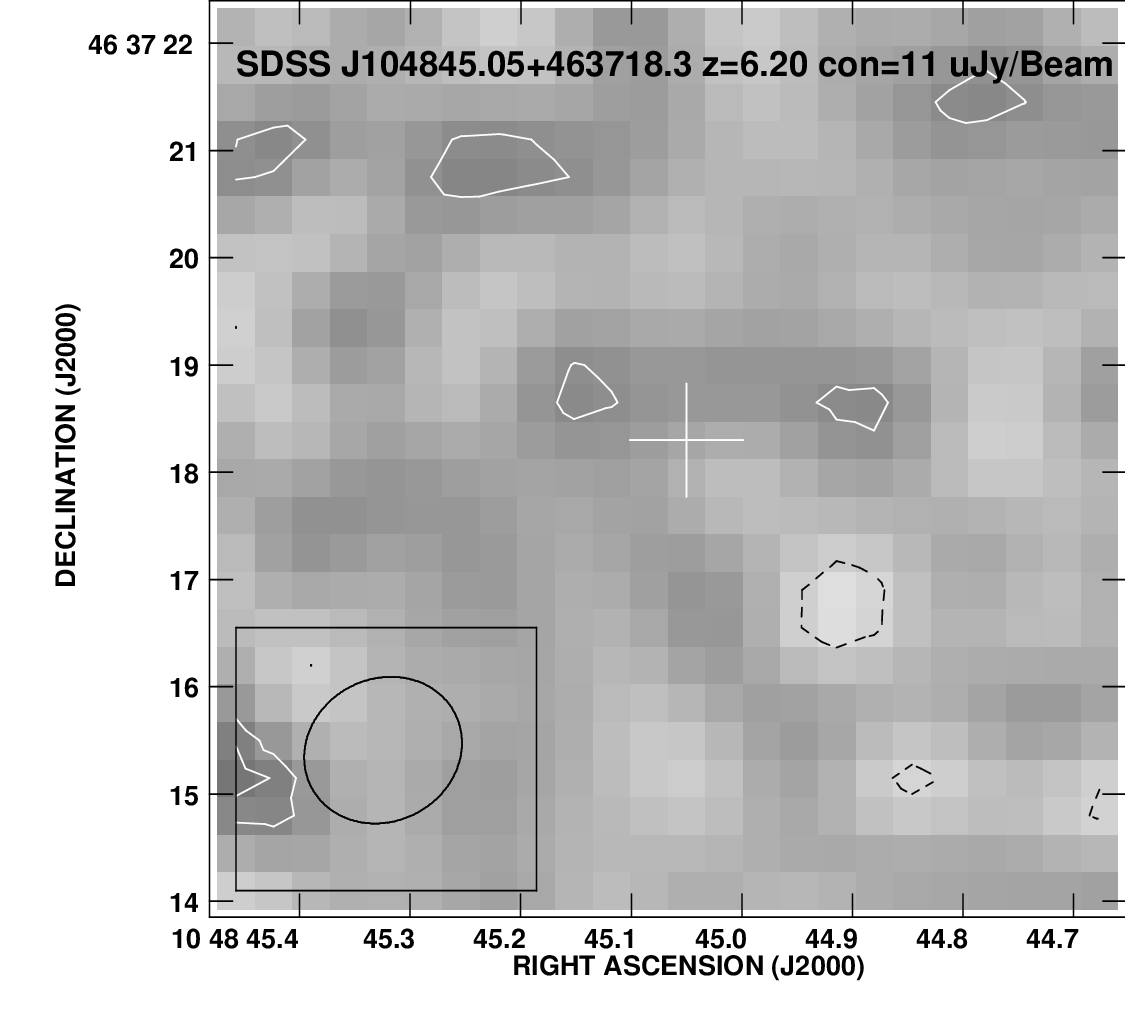}{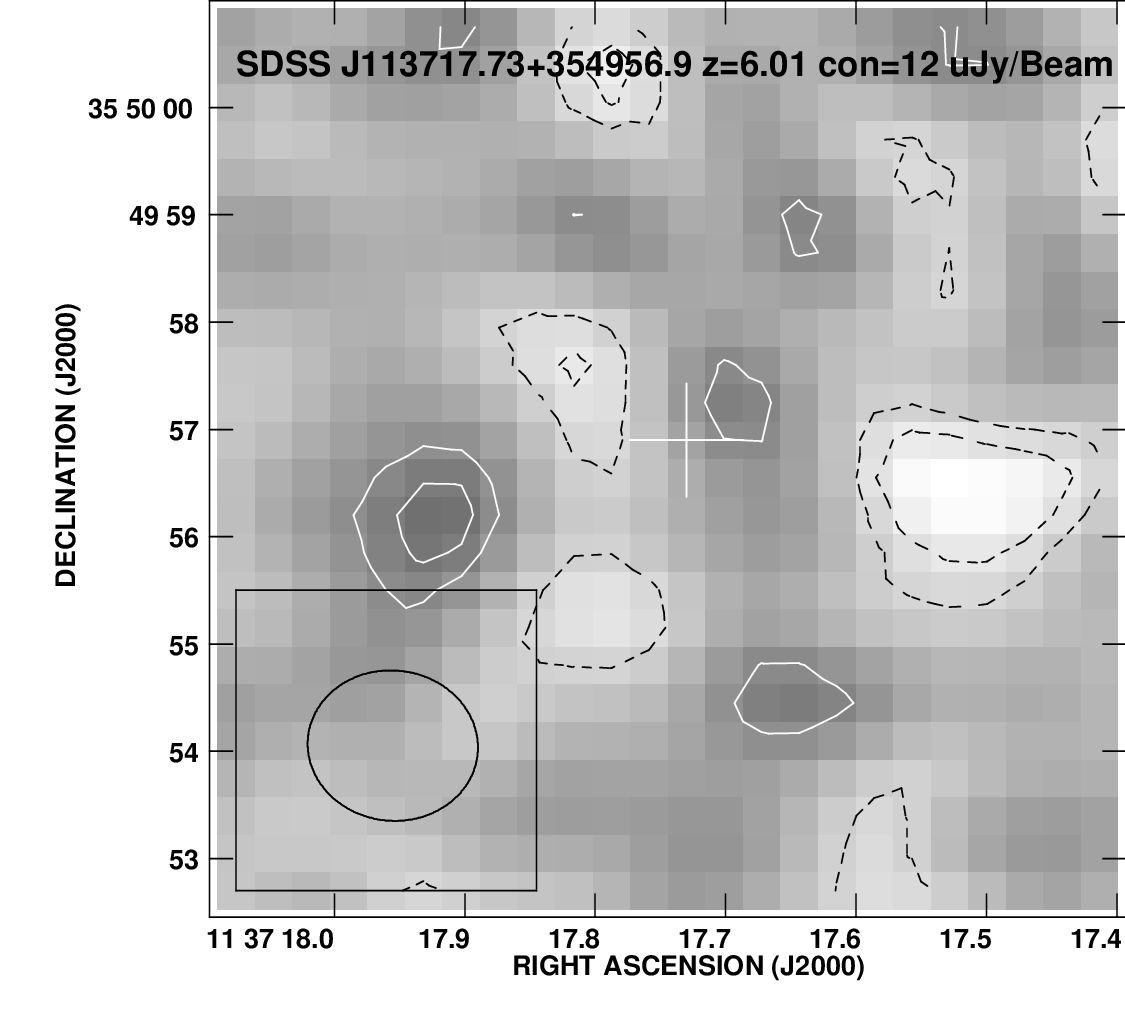}\\
\plottwo{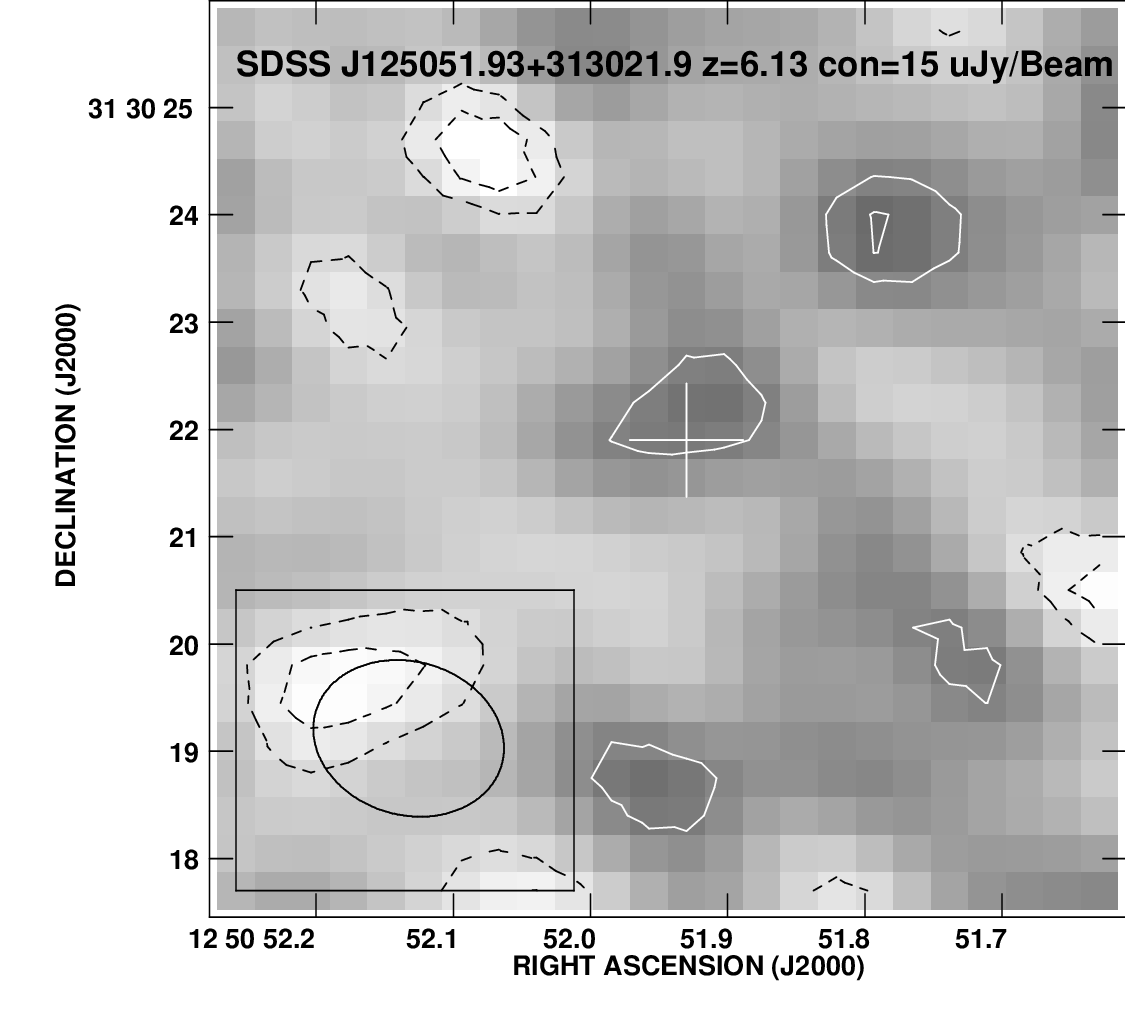}{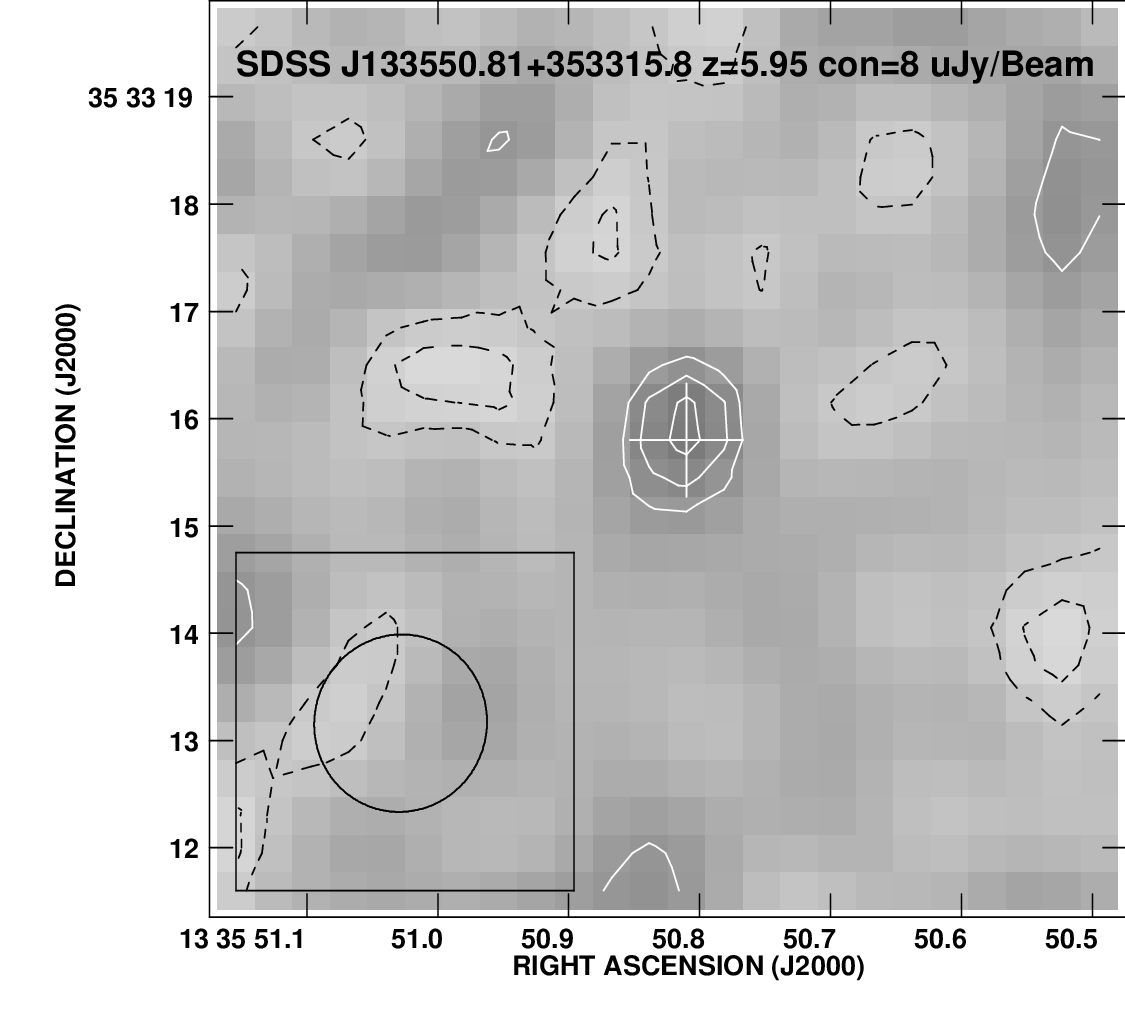}\\
Fig. 1 -- Continued
\end{figure}

\begin{figure}
\figurenum{1}
\plottwo{f1i.eps}{f1j.eps}\\
\plottwo{f1k.eps}{f1l.eps}\\
Fig. 1 -- Continued
\end{figure}


\begin{figure}
\figurenum{2}
\epsscale{0.6} \plotone{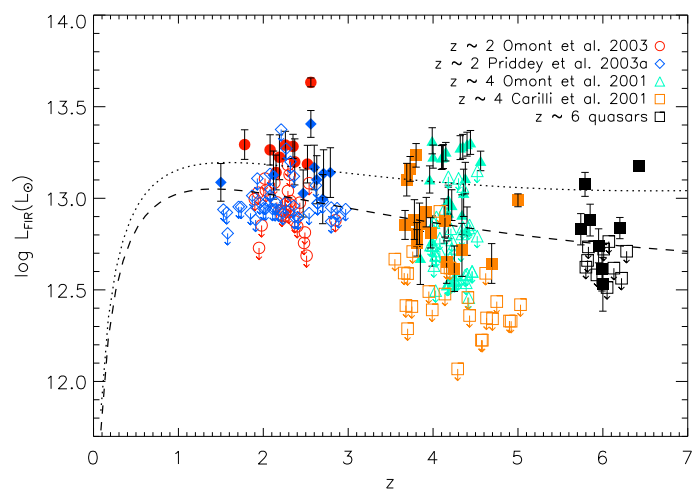}\\
 \figcaption{The logarithm of the FIR luminosity ($\rm log
L_{FIR}$) versus redshift for different samples. The filled
symbols represent detections with 1$\rm \sigma$ errors. The open
symbols with arrows denote upper limits: we adopt the larger value
of either the 2$\rm \sigma$ rms or the measured value at the optical
position plus 1$\rm \sigma$ rms as upper limits for non-detected
sources. The dashed and dotted lines represent the typical
$\rm 3\sigma$ detection limits of MAMBO at 250 GHz and SCUBA at 350
GHz, respectively, namely $\rm S_{250GHz}=2.4\,mJy$ (this work) and
$\rm S_{350GHz}=10\,mJy$ (Priddey et al. 2003a). \label{fig2}}
\end{figure}


\begin{figure}
\figurenum{3}
\epsscale{0.6}
\plotone{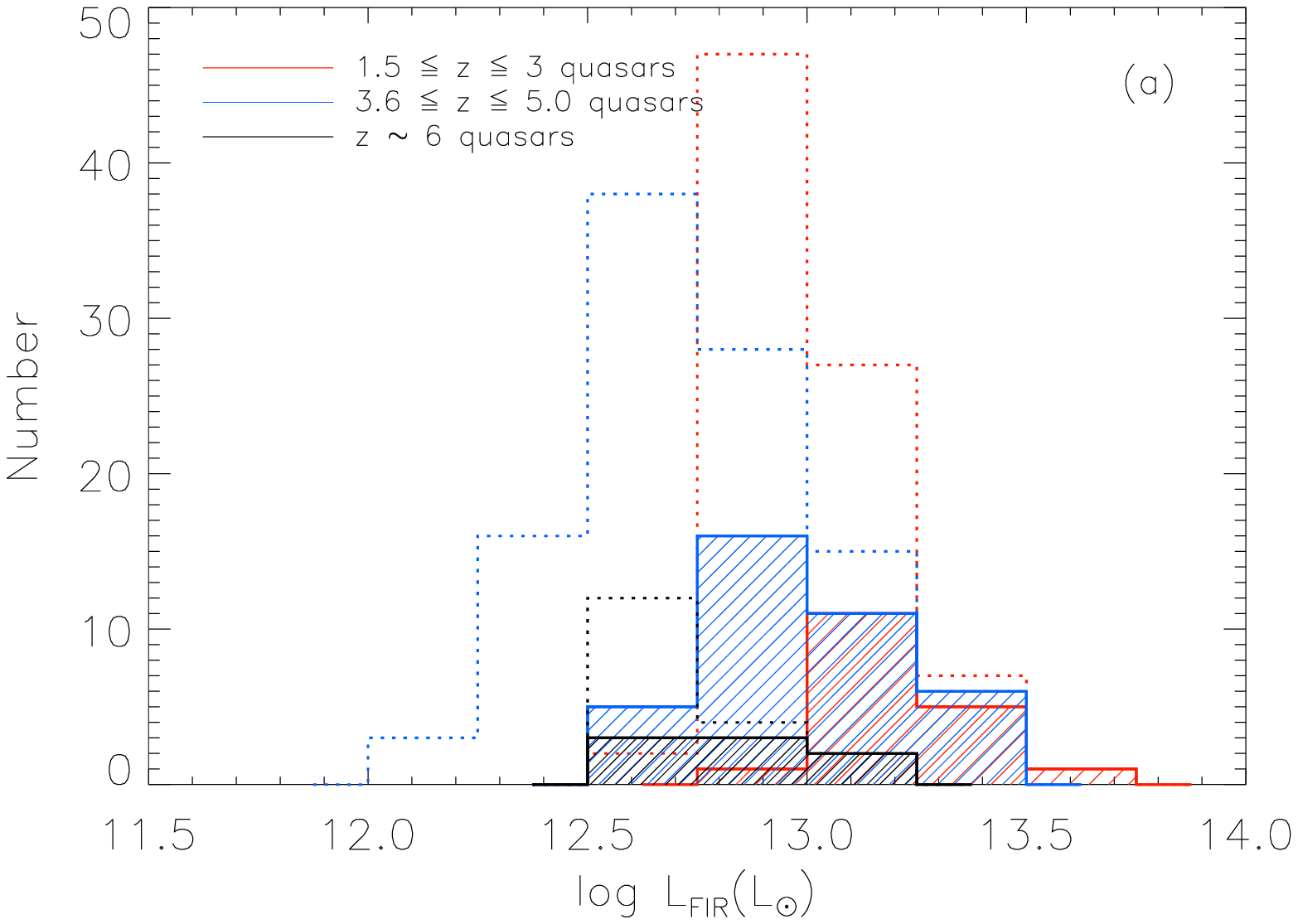}\\
\plotone{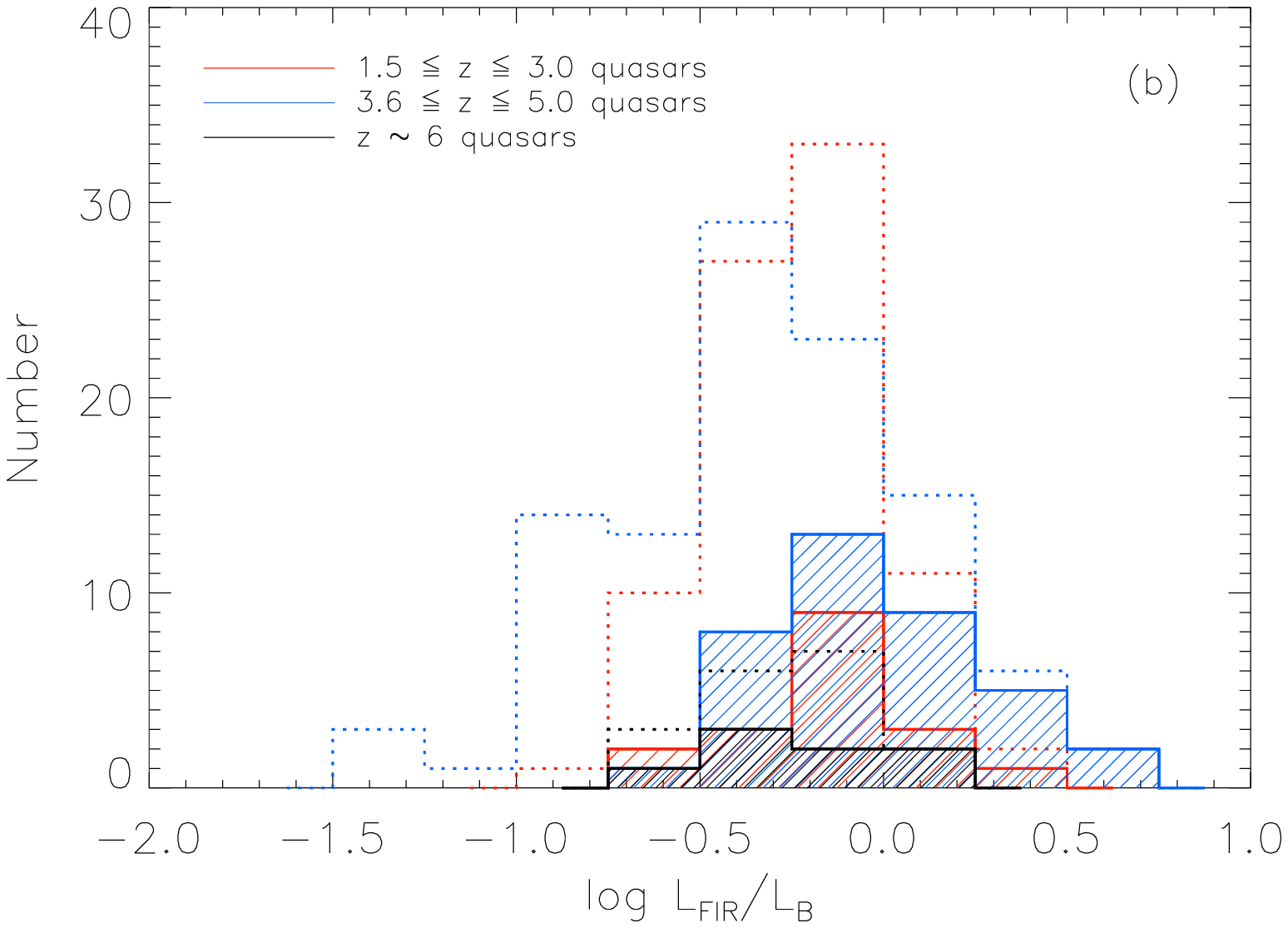}\\
\figcaption{(a) The distribution of the logarithm of the FIR
luminosity for quasars at different redshifts. The z$\sim$6 quasars
are plotted as black lines. The $\rm 1.5\leq z\leq 3$ group (red
lines) is combined with the samples from Omont et al. (2003) and
Priddey et al. (2003a), while the $\rm 3.6\leq z \leq 5$ group (blue
lines) is from Omont et al. (2001) and Carilli et al. (2001). For
all the samples, the shaded areas represent detections and dotted
lines denote upper limits: the upper limits are calculated as in
Figure 2. (b) The distribution of the logarithm of the FIR-optical
ratio ($\rm log L_{FIR}/L_{B}$) for quasars at different redshifts.
\label{fig3}}
\end{figure}

\begin{figure}
\figurenum{4}
\epsscale{0.8} \plotone{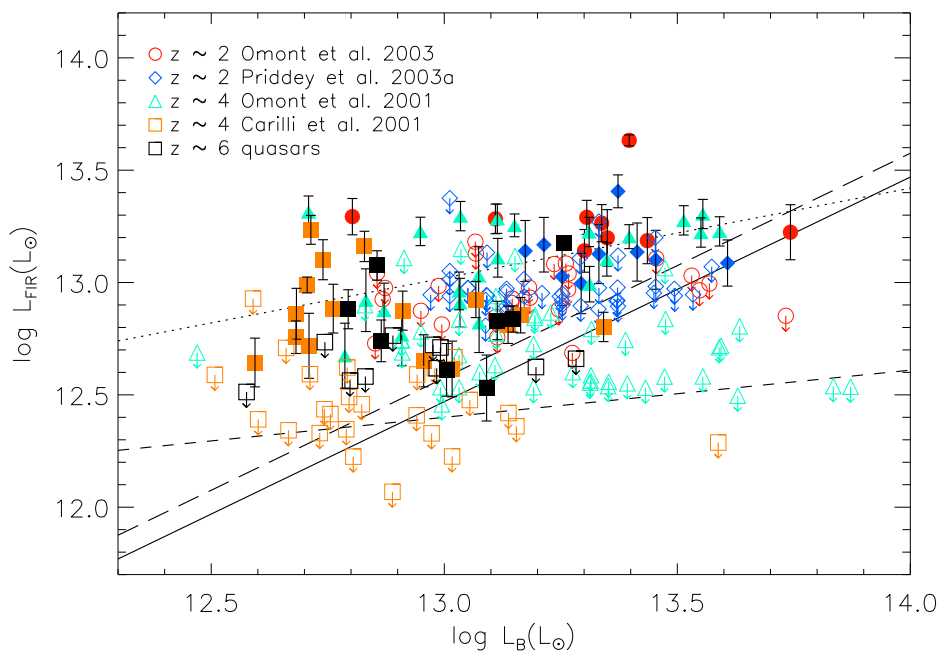} \figcaption{The logarithm of the FIR
luminosity is plotted against the logarithm of the B band luminosity
for all the (sub)mm observed quasars. The symbols are the same as in
Figure 2. The dashed line is the linear regression result for the
(sub)mm observed samples at all redshifts including upper limits,
while the dotted line is the result using the detected objects only.
The solid and long dashed lines represent the FIR-to-B band luminosity 
ratios of local radio quiet and radio loud quasar templates 
in Elvis et al. (1994), respectively.
\label{fig4} }
\end{figure}


\begin{figure}
\figurenum{5}
\epsscale{0.8} \plotone{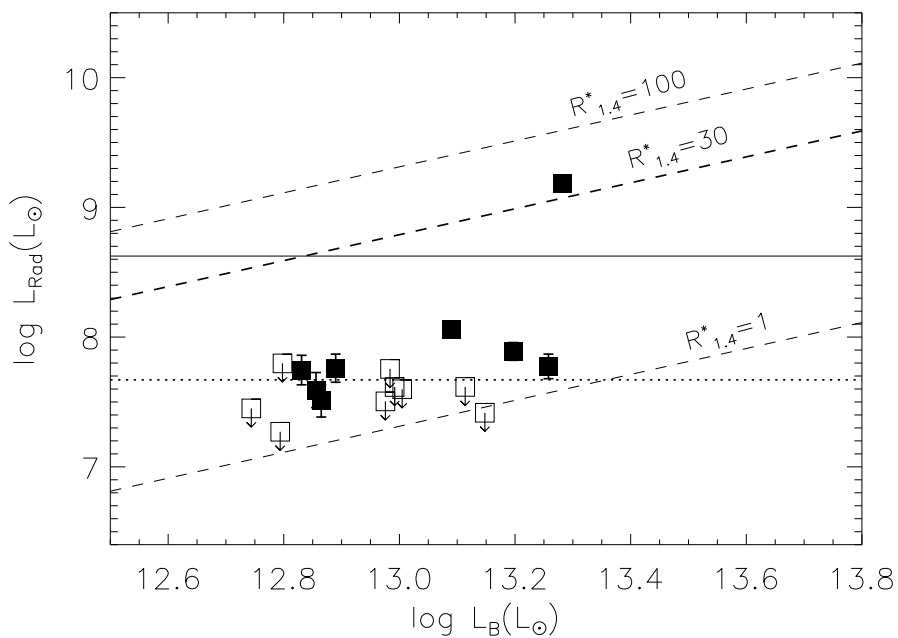} \figcaption{The logarithm of the
rest frame 1.4GHz radio luminosity ($\rm log L_{Rad}$) is plotted
against the logarithm of the B band luminosity ($\rm log L_{B}$) for
the z$\sim$6 quasars observed with the VLA at 1.4GHz. The filled
squares represent detections with 1$\rm \sigma$ errors, and the open
squares with arrows denote upper limits: the upper limits are
calculated as in Figure 2. The dashed lines represent rest frame
radio-to-optical ratios ($\rm R^{*}_{1.4}$) of 100, 30 (separation
of radio loud and quiet), and 1. The
dotted line denotes the typical 3$\rm \sigma$ detection limits of
our VLA observations, while the solid line shows the 3$\rm \sigma$
detection limit of the FIRST survey. \label{fig5}}
\end{figure}

\begin{figure}
\figurenum{6}
\epsscale{0.8} \plotone{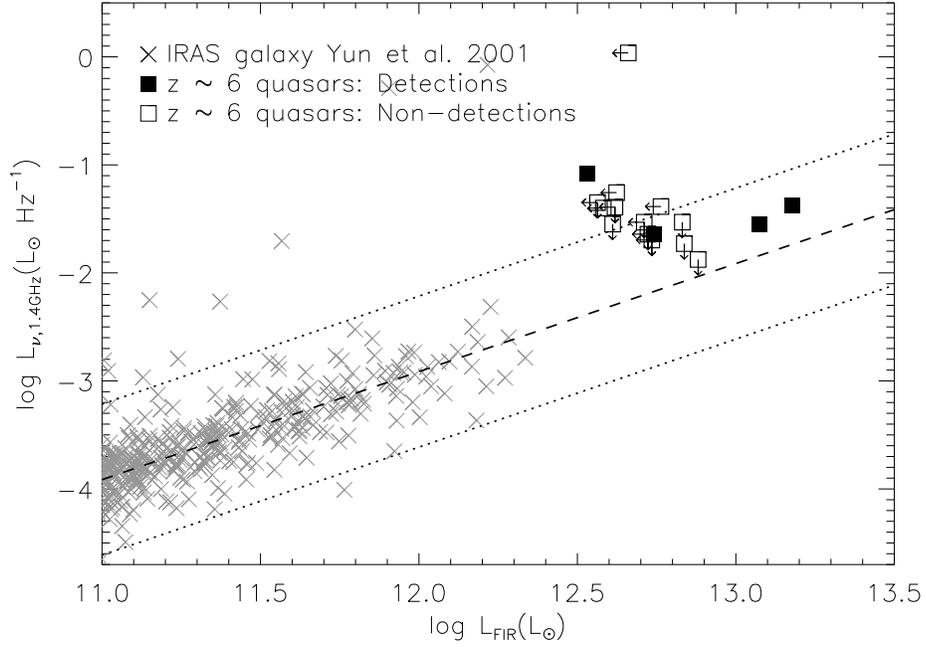} \figcaption{The logarithm of the
rest frame 1.4 GHz radio luminosity density ($\rm log L_{\nu
,1.4GHz}$) versus the logarithm of the FIR luminosity ($\rm log
L_{FIR}$). The filled squares represent the detections in both radio
and (sub)mm bands, and the open squares with arrows represent the
upper limits of the non-detections in either radio or (sub)mm. The
upper limits are calculated as in Figure 2. The crosses represent
the IRAS 2Jy sample of galaxies in Yun et al. (2001) and the dashed
line indicates the typical radio-to-FIR correlation in star forming
galaxies, with correlation parameter $\rm q=2.34$ (Yun et al. 2001).
The dotted lines represent excesses five times above and below the
typical $\rm q$ value.\label{fig6}}
\end{figure}

\vspace{-5em}
\begin{center}
\begin{table}
{\scriptsize \caption{1.4GHz and 250GHz results of 13
z$\sim$6 quasars. \label{tbl-1}}
\begin{tabular}{lccccccr}
\hline \noalign{\smallskip}
SDSS name & z &$\rm m_{1450\AA}$& \multicolumn{4}{c}{radio detection at 1.4 GHz}&$\rm S_{250GHz}$\\
\cline{4-7}
          &   & &$\rm S_{1.4opt}$\tablenotemark{b}&$\rm RA_{peak}$\tablenotemark{c}&$\rm Dec_{peak}$\tablenotemark{c}&$\rm S_{peak}$\tablenotemark{c}& \\
 & & &$\mu$Jy\,Beam$^{-1}$&h m s&\ $^{\circ}$\ $^{'}$\ $^{''}$&$\rm \mu$Jy\,Beam$^{-1}$&mJy\\
(1)&(2)&(3)&(4)&(5)&(6)&(7)&(8)\\
\noalign{\smallskip} \hline \noalign{\smallskip}
J000239.39+255034.8&5.80&19.02 &81$\pm$14&00:02:39.36&25:50:34.9&{\bf 89}\tablenotemark{d}&0.20$\pm$0.88\\
J000552.34$-$000655.8&5.85&20.03 &40$\pm$130& & & &\\
J081827.40+172251.8&6.00&19.34 &119$\pm$12&08:18:27.39&17:22:51.8&{\bf 123}&{\bf 1.45$\pm$0.49}\\
J084035.09+562419.9&5.85& 20.04&12$\pm$9& & & &{\bf 3.20$\pm$0.64}\\
J092721.82+200123.7&5.79&19.87 &33$\pm$14&09:27:21.82&20:01:23.1&{\bf 45}&{\bf 4.98$\pm$0.75}\\
J104845.05+463718.3&6.20&19.25 &6$\pm$11& & & &{\bf 3.00$\pm$0.40}\tablenotemark{a}\\
J113717.73+354956.9&6.01&19.63 &9$\pm$17& & & &0.10$\pm$1.13\\
J125051.93+313021.9&6.13&19.64 &37$\pm$21& & & &0.07$\pm$0.90\\
J133550.81+353315.8&5.95& 19.89&35$\pm$10&13:35:50.81&35:33:15.9&{\bf 35}&{\bf 2.34$\pm$0.50}\\
J141111.29+121737.4&5.93&19.97 &44$\pm$16&14:11:11.29&12:17:38.0&{\bf 61}&1.00$\pm$0.62\\
J143611.74+500706.9&5.83&20.16&6$\pm$16& & &  &-0.21$\pm$1.14\\
J160253.98+422824.9&6.07&19.86 &53$\pm$15&16:02:53.95 &42:28:24.9 &{\bf 60}&1.82$\pm$0.86\\
J162331.81+311200.5&6.22&20.13 &24$\pm$31& & &  &0.17$\pm$0.80\\
\noalign{\smallskip} \hline
\end{tabular}
}\tablenotetext{a}{Result taken from Bertoldi et al. (2003a).}
\tablenotetext{b}{Surface brightness at the optical quasar position
plus the $\rm 1\sigma$ rms on the radio map.}
\tablenotetext{c}{Position and surface brightness of the detected
radio peak.}\tablenotetext{d}{The detections are marked as
boldface.}
\end{table}
\end{center}


\begin{table}
\begin{center}
{\scriptsize
\caption{Published results from the literature.
\label{tbl-2}}
\begin{tabular}{rcccccl}
\hline \noalign{\smallskip}
SDSS name&z&$\rm m_{1450\AA}$&$\rm S_{250GHz}$&$\rm S_{350GHz}$&$\rm S_{667GHz}$&$\rm S_{1.4GHz}$\\
         & &          & mJy & mJy & mJy & uJy\\
(1)&(2)&(3)&(4)&(5)&(6)&(7)\\
\noalign{\smallskip} \hline \noalign{\smallskip}
J083643.85+005453.3&5.82&18.81&-0.4$\pm$1.0\tablenotemark{a}&1.7$\pm$1.5\tablenotemark{b}&-24$\pm$10\tablenotemark{b}&{\bf 1740$\pm$40}\tablenotemark{f,g}\\
J103027.10+052455.0&6.28&19.66&-1.1$\pm$1.1\tablenotemark{a}&1.3$\pm$1.0\tablenotemark{b}&-21$\pm$10\tablenotemark{b}&-3$\pm$20\tablenotemark{a}\\
J104433.04$-$012502.2&5.74&19.21&-&{\bf 6.1$\pm$1.2}\tablenotemark{b}&-&-15$\pm$24\tablenotemark{a}\\
J114816.64+525150.3&6.42&19.03&{\bf 5$\pm$0.6}\tablenotemark{c}&{\bf 7.8$\pm$0.7}\tablenotemark{d}&{\bf 24.7$\pm$7.4}\tablenotemark{d}&{\bf 55$\pm$12}\tablenotemark{e}\\
J130608.26+035626.3&5.99&19.55&-1.0$\pm$1.0\tablenotemark{a}&{\bf 3.7$\pm$1.0}\tablenotemark{b}&-7$\pm$14\tablenotemark{b}&14$\pm$21\tablenotemark{a}\\
J163033.90+401209.6&6.05&20.64&0.8$\pm$0.6\tablenotemark{c}&2.7$\pm$1.9\tablenotemark{d}&15.4$\pm$9.6\tablenotemark{d}&$<440$\tablenotemark{c}\\
\noalign{\smallskip} \hline
\end{tabular}
}\tablenotetext{a}{Petric et al. (2003).} \tablenotetext{b}{Priddey
et al. (2003b).} \tablenotetext{c}{Bertoldi et al. (2003a).}
\tablenotetext{d}{Robson et al. (2004).} \tablenotetext{e}{Carilli
et al. (2004).}\tablenotetext{f}{weighted average of the results in
Petric et al. (2003) and the FIRST survey.}\tablenotetext{g}{The
detections are marked as boldface.}
\end{center}
\end{table}


\begin{table}
\begin{center}
{\scriptsize \caption{Correlation tests \label{tbl-5}}
\begin{tabular}{rccc}
\hline \noalign{\smallskip}
 Group & Number & Kendall's tau  &$\rm P_{null}$ \\
  (1)       & (2)    & (3)   &(4)      \\
\noalign{\smallskip} \hline \noalign{\smallskip}
  z$\sim$6& 18 & 0.2222 &  0.4402 \\
  Total   & 208& 0.1246 &  0.0575\\
Detection     & 64 & 0.6310 &  0.0002\\
\noalign{\smallskip} \hline
\end{tabular}
}
\end{center}
\end{table}


\begin{table}
\begin{center}
{\scriptsize \caption{Derived parameters of the z$\sim$6 quasars
\label{tbl-3}}
\begin{tabular}{rccccc}
\hline \noalign{\smallskip}
SDSS name & log$L_{B}$ & log$\rm L_{FIR}$ & log$\rm L_{Rad}$ & $\rm R^{*}_{1.4}$ & $\rm q$ \\
          & ($\rm L_{\odot}$)&($\rm L_{\odot}$)&($\rm L_{\odot}$)& & \\
(1)&(2)&(3)&(4)&(5)&(6)\\
\noalign{\smallskip} \hline \noalign{\smallskip}
J000239.39+255034.8& 13.20& $<$12.62&7.89&2.40& $<$1.31\\
J081827.40+172251.8& 13.09&    12.53&8.06&4.55& 1.04 \\
J083643.85+005453.3& 13.28& $<$12.66&9.18&38.68& $<$0.05\\
J084035.09+562419.9& 12.79&    12.88&$<$7.27&$<$1.46&$>$2.18\\
J092721.82+200123.7& 12.86&    13.08& 7.59& 2.65&2.06\\
J103027.10+052455.0& 12.99& $<$12.71&$<$7.62&$<$2.05&--\\
J104433.04-012502.2& 13.11&    12.83&$<$7.62&$<$1.55&$>$1.79\\
J104845.05+463718.3& 13.15&    12.84&$<$7.42&$<$0.90&$>$1.99\\
J113717.73+354956.9& 12.98& $<$12.72&$<$7.50&$<$1.65&--\\
J114816.64+525150.2& 13.26&    13.18&7.77&1.60&1.98\\
J125051.93+313021.9& 12.98& $<$12.62&$<$7.75&$<$2.87&--\\
J130608.26+035626.3& 13.00&    12.61&$<$7.60&$<$1.91&$>$1.59\\
J133550.81+353315.8& 12.86&    12.74&7.51&2.14&1.80\\
J141111.29+121737.4& 12.83& $<$12.58&7.75&4.01&$<$1.41\\
J143611.74+500706.9& 12.74& $<$12.73&$<$7.45&$<$2.48&--\\
J160253.98+422824.9& 12.89& $<$12.76&7.76&3.62&$<$1.58\\
J162331.81+311200.5& 12.80& $<$12.56&$<$7.80&$<$4.86&--\\
J163033.90+401209.6& 12.58& $<$12.51& -- & -- &--\\
\noalign{\smallskip} \hline
\end{tabular}\\
}
\end{center}
Note. --- Col. (1), the name of the source, Col. (2), the B band
luminosity (see the footnote in Section 2 for the calculation), Col.
(3), the rest frame FIR luminosity, Col. (4), the rest frame 1.4 GHz
radio luminosity, Col. (5), the radio-optical ratio $\rm
R^{*}_{1.4}$, and Col. (6) the FIR-to-radio correlation parameter
$\rm q$.
\end{table}

\end{document}